\documentclass[%
 preprint,
superscriptaddress,
 amsmath,amssymb,
 aps,
 prb,
floatfix,
]{revtex4-2}

\usepackage{graphicx}
\usepackage{dcolumn}
\usepackage{bm}
\usepackage[utf8]{inputenc}
\usepackage[T1]{fontenc}
\usepackage{mathptmx}
\usepackage{textcomp}
\usepackage{siunitx}
\usepackage{bigints}

\usepackage{hyperref}
\usepackage[switch]{lineno}
\usepackage[table,xcdraw,dvipsnames]{xcolor}
\usepackage{subcaption}
\usepackage{booktabs}
\usepackage{float}
\usepackage{color}
\usepackage{ulem}
\usepackage{multirow}

\begin{document}

\preprint{APS/123-QED}

\title{Generation and annihilation of skyrmions and antiskyrmions in magnetic heterostructures}

\author{Sabri~Koraltan}
 \affiliation{Faculty of Physics, University of Vienna, Kolingasse 14-16, A-1090, Vienna, Austria}
\author{Claas~Abert}%
\affiliation{Faculty of Physics, University of Vienna, Kolingasse 14-16, A-1090, Vienna, Austria}
\affiliation{Research Platform MMM Mathematics - Magnetism - Materials, University of Vienna, Vienna 1090, Austria}
\author{Florian~Bruckner}%
\affiliation{Faculty of Physics, University of Vienna, Kolingasse 14-16, A-1090, Vienna, Austria}
\author{Michael~Heigl}%
\affiliation{Institute of Physics, University of Augsburg, Augsburg 86159, Germany}
\author{Manfred~Albrecht}%
\affiliation{Institute of Physics, University of Augsburg, Augsburg 86159, Germany}
\author{Dieter~Suess}
\affiliation{Faculty of Physics, University of Vienna, Kolingasse 14-16, A-1090, Vienna, Austria}
\date{\today}
\affiliation{Research Platform MMM Mathematics - Magnetism - Materials, University of Vienna, Vienna 1090, Austria}             

\begin{abstract}

We demonstrate the controlled generation and annihilation of (anti)skyrmions with tunable chirality in magnetic heterostructures by means of micromagnetic simulations. By making use of magnetic (anti)vortices in patterned ferromagnetic layer, we stabilize full lattices of (anti)skyrmions in an underlying skyrmionic thin film in a reproducible manner. The stability of the (anti)skyrmion depends on the polarization of the (anti)vortex, whereas their chirality is given by those of the (anti)vortices. Furthermore, we demonstrate that the core coupling between the (anti)vortices and (anti)skyrmions allows to annihilate the spin-objects in a controlled fashion by applying short pulses of in-plane external magnetic fields, representing a new key paradigm in skyrmionic devices.

\end{abstract}

\maketitle


\section{Introduction}
\label{sec:introduction}
Magnetic quasi-particles with non-trivial magnetization states have been researched intensively since the first discovery of skyrmions \cite{bogdanov_thermodynamically_1994, bogdanov_thermodynamically_1994}. The skyrmions, local whirls in out-of-plane magnetized systems, are topologically protected spin textures, usually stabilized by inversion symmetry breaking energies such as Dzyaloshinskii-Moriya interaction~\cite{boulle_room-temperature_2016, everschor-sitte_perspective_2018, mandru_coexistence_2020, jena_observation_2020, jena_elliptical_2020}. Recently, it was shown that Bloch skyrmions can become stable spin-textures in systems dominated by purely dipolar interactions, by making use of multilayered systems such as Iron-Gadolinium based multilayers\cite{montoya_tailoring_2017, montoya_resonant_2017}.

Visionary ideas to create skyrmionic devices were supposed to make use of the increased stability attributed to the topological protection. The first skyrmion race track was proposed in such a fashion, that one could store the information in the distance in between two skyrmions~\cite{fert_skyrmions_2013}, where basically the presence of one skyrmion would be equivalent to the bit \texttt{1}, while its absence would represent the bit \texttt{0}. However, the non-constant velocities of the individual skyrmions, the skrmion hall-effect, as well as the pinning sites in the racetracks have shown that this principle of storage is more challenging~\cite{suess_repulsive_2018, suess_spin_2019}.

With the discovery of the antiskyrmions~\cite{nayak_magnetic_2017, jena_elliptical_2020}, the topological counterpart of the skyrmion, a new storage principle arises. Namely, the skyrmion could store the bit \texttt{0}, whereas the antiskyrmion could store the bit \texttt{1}. The recent discovery of mainly dipolar stabilized antiskyrmions~\cite{heigl_dipolar-stabilized_2021}, and the coexistence of different spin-textures in one material opens a new paradigm in the skyrmionic storage devices~\cite{mandru_coexistence_2020, everschor-sitte_perspective_2018}. However the random distribution of the spin objects, and the lack of controlability of nucleation sites, raises new problems.

Tuning and controlling the macroscopic properties of a magnetic material starts to become a very important aspect today~\cite{chumak_advances_2022, Back_roadmap_2020}, while the field of magnetic meta-materials and heterostructures attracts more and more interest~\cite{skjaervo_advances_2020, skovdal_temperature_2021, Papp_inverse_2021, Wang_inverse_2021}. Placing ordered arrays of ferromagnetic nanostructures on top of the host material, allows to control different properties of these materials, as recently shown in superconducting and ferromagnetic heterostructures, as well as in inverse-designed~\cite{Papp_inverse_2021} and regular magnonic crystals~\cite{Chumak_magnonic_2017, chumak_advances_2022}. The simple working principle of meta-materials can be expressed by considering a host material, where a certain property has to be tuned, and it is truly challenging to accomplish this by conventional matters. A second system, which has been previously extensively researched, and which is very well controllable, is coupled to the host system. By easily tuning the properties of the meta-material, one tries to achieve a desired macroscopic state in the host material.

In this work, we present a micromagnetic study where we use a skyrmionic host material with magnetic properties similar to Fe/Ir/Gd multilayers, and couple it to nanodot or interconnected square lattices in order to nucleate spin textures in the host system. While the former gives rise to a skyrmion lattice, the latter can be used to nucleate antiskyrmions in a controlled manner. Due to low magnetic moment of the underlayer the coupling of both layers is dominated by exchange. The permalloy based meta-materials can harbor magnetic vortex and anti-vortex structures, which lead to a core coupling between the two layers. We demonstrate that the spin-textures can be annihilated easily when picosecond ranged pulses of in-plane magnetic fields are applied. 

This paper is organized as follows. The geometrical modeling of the lattices, as well as details about the micromagnetic simulations are presented in Sec.~\ref{sec:modeling}. The findings are presented and discussed in Sec.~\ref{sec:results}, where we give a deeper insight into the coupling and relaxation mechanisms of the spin-textures, as well as present the topologically non-violent erasing of (anti) skyrmions. We summarize and conclude our main findings in Sec.~\ref{sec:conclusion}.

\section{Modeling}
\label{sec:modeling}
In this work, we make use of micromagnetic simulations, to describe the coupling between a vortex (VL) and skyrmionic layer (SkL). For this purpose we are using the finite differences based micromagnetic simulation software \texttt{magnum.af}\cite{magnum_af}, and a hybrid finite and boundary element method software \texttt{magnum.fe}.

The ferrimagnetic SkL, which in reality is a multilayered material, is modeled as a ferromagnet with the low saturation magnetization $M_s^{\mathrm{SkL}} = \SI{225}{kA/m}$, the perpendicular uniaxial anisotropy constant $K_u^{\mathrm{SkL}} = \SI{22.4}{kJ/m^3}$, and the exchange stiffness constant $A_{ex}^{\mathrm{SkL}} = \SI{6}{pJ/m}$~\cite{heigl_dipolar-stabilized_2021}.

For the meta-material in the performed numerical experiments, we choose Permalloy with $M_s^{\mathrm{VL}} = \SI{800}{kA/m}$, $K_u^{\mathrm{VL}} = \SI{0}{kJ/m^3}$, $A_{ex}^{\mathrm{VL}} = \SI{13}{pJ/m}$.

We include the energetic contributions from the demagnetization, anisotropy, exchange, and external fields into the effective field term, which is then used to solve the Landau-Lifshitz-Gilbert equation at moderate damping parameter $\alpha=0.2$. Note that the SkL and VL are purely exchange and stray field coupled. Additional interlayer couplings, e.g. RKKY coupling similar to artificial skyrmion lattices~\cite{sun_creating_2013, gilbert_realization_2015}, are not used.

The skrymions and antiskyrmions differ from each other by their opposite signed topological charge
\begin{equation}
N_{\mathrm{sk}} = \bigintss \dfrac{1}{4\pi}\textbf{\textit{m}}\cdot \left(\dfrac{\partial \textbf{\textit{m}}}{\partial x}\times \dfrac{\partial \textbf{\textit{m}}}{\partial y} \right)\mathrm{d}x\mathrm{d}y,
\label{eq:nsk}
\end{equation}
where $\boldsymbol{m}$ is the normalized magnetization vector. We calculate the topological charge to show the temporal evolution of the skyrmions.

For our micromagnetic investigations we consider two unit geometries to build the (anti)vortex layer: the nanodisk and the nanocross.
Lattices built using these components have been extensively researched over the past years and are very well understood. While former can host magnetic vortices in an ordered fashion, the latter reassembles an interconnected artificial spin ice lattice, where different microscopic magnetic configurations can be chosen. While a low magnetic moment of the skyrmionic layer is crucial to stabilize the (anti)skyrmions, the low strayfield coupling between the two layers leads to stable (anti)vortices. Hence, the coupling between the host and meta-material is exchange dominated.

In the following we discuss briefly the geometrical description of the used hetero-structures.

\subsection{Nanodot lattice}
\label{sec:modelingNL}

Without loss of generality, magnetic nanodisks under certain sizes have at least two accessible ground states: the uniform magnetization and the magnetic vortex. The latter can be used for sensing magnetic fields~\cite{suess_topoligically_2018} with a large linear range, emission of very short-wavelength spin-waves~\cite{wintz_magnetic_2016, sluka_emission_2019}, or recently, also for reservoir computing~\cite{gartside_reconfigurable_2022}.

When one arranges the magnetic vortices in a 2D array, their magnetic configuration can become unstable~\cite{skovdal_temperature_2021}. For their increased meta-stability one requires well defined disk dimensions, above the critical stability threshold. Commonly, thicknesses between $30$ and $\SI{100}{nm}$ are chosen, while keeping the lateral dimension between $500$ and $\SI{1000}{nm}$. We remind here that the low stray-field coupling is essential to obtain meta-stable magnetic vortices.

\begin{figure}
    \centering
    \includegraphics[width=\columnwidth]{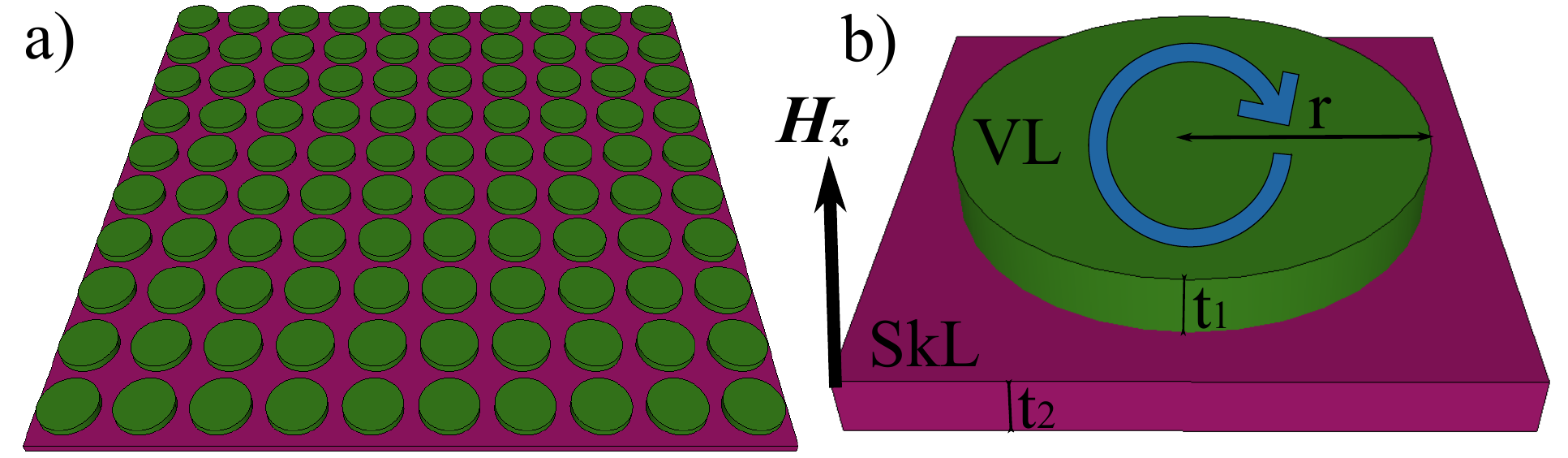}
    \caption{Schematic illustration of the vortex~(green) and skyrmion~(pink) layers, where (a) shows the entire lattice, and (b) shows one isolated nanodisk on top of the host material.}
    \label{fig:fig01}
\end{figure}

Figure~\ref{fig:fig01}a) shows a schematic illustration of the considered vortex lattice as the meta-material. The radius of each nanodisk is $r = \SI{200}{nm}$ as shown in Fig.~\ref{fig:fig01}b), while the thickness is $t=\SI{30}{nm}$.

The total SkL has the dimensons $5000\times5000\times\SI{40}{nm^3}$, and we arrange $100$ nanodisks on top with a gap of \SI{100}{nm} between each other. In Sec.~\ref{sec:results} we will only illustrate a small portion of the lattice to keep the figures well visible.

\subsection{Square Lattice}
\label{sec:modelingSL}

With the goal to stabilize an antiskyrmion lattice we consider an interconnected square ice lattice which is similar to an aritifical spin ice lattice~(ASI) \cite{wang_artificial_2006, skjaervo_advances_2020, koraltan_dependence_2020, hofhuis_thermally_2020,farhan_direct_2013}. Commonly, ASIs  are frustrated lattices consisting of lithographically patterned, stadium-shaped, and thermally active magnetic nanoislands. However, interconnected lattices were investigated intensively, as they share similar frustration properties.
The ground state of the ASI vertices is given by the so called $\textit{ice-rule}$, where the magnetizations of two nano-islands (or arms, like in our lattice) points to the center of the vertex while the other two are oriented out. The long-range-ordered ground state is achieved with an antiferromagnetic arrangement within the $\textit{ice-rule}$, usually referred to skyrmionsas the \textit{Type-I} configuration.

\begin{figure}
    \centering
    \includegraphics[width=\columnwidth]{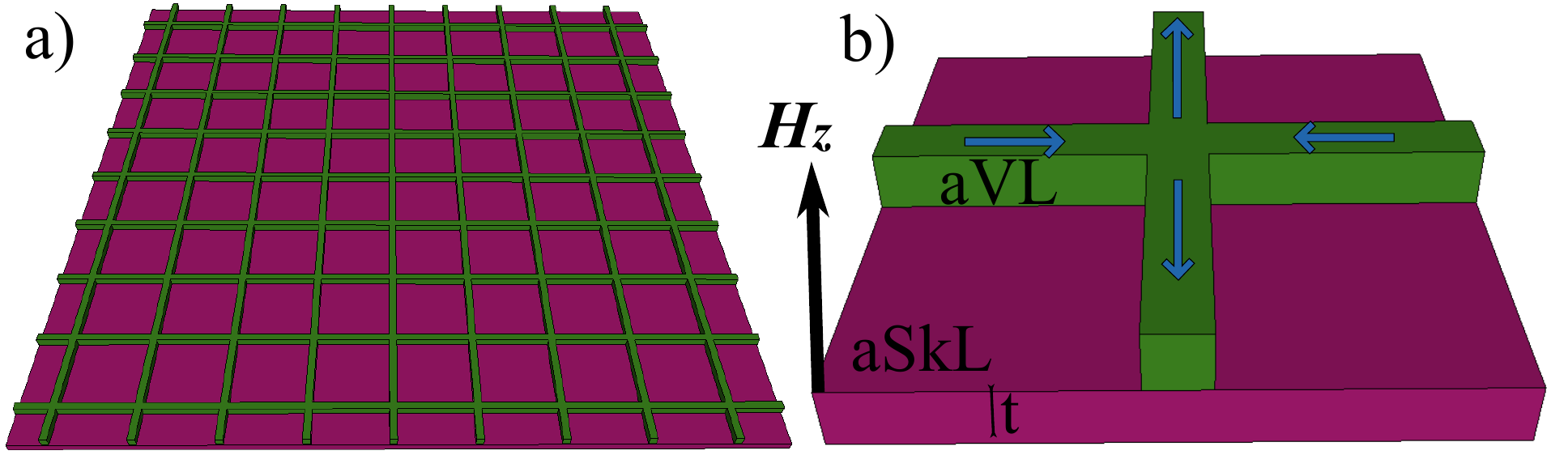}
    \caption{Square lattice (green), as the magnetic meta-material, exchange coupled to the skyrmionic layer (pink) is illustrated in (a), while (b) visualized a single nanocross with the \textit{ice-rule} ground state depicted by blue arrows.}
    \label{fig:fig02}
\end{figure}

Comparing this configuration to the stabilization mechanisms of antivortices~\cite{gliga_switching_2008}, one sees that one can obtain an antivortex at the center of the cross~(green), shown in Fig.~\ref{fig:fig02}a), where the magnetization obeys the \textit{ice-rule} and represents the ground state. Using the same SkL~(pink) as in Fig.~\ref{fig:fig01}, we build a square lattice and a ferrimagnet heterostructure which can host antivortices and antiskyrmions individually.   

\section{Results}
\label{sec:results}
\subsection{Vortex skyrmion coupling}
\label{sec:resultsSKL}
Using the finite differences based simulation software \texttt{magnum.af}\cite{magnum_af} we magnetize randomly each cell of the magnetic geometry schematically shown in Fig.~\ref{fig:fig01}a). First, we relax the system by numerically solving the LLG at vanishing external fields. Magnetic vortices with clockwise and counterclockwise chirality, and positive or negative vortex core polarity form in the VL, and small domains and stripe domains in the SkL.

\begin{figure*}
	\centering
	\includegraphics[width=\textwidth]{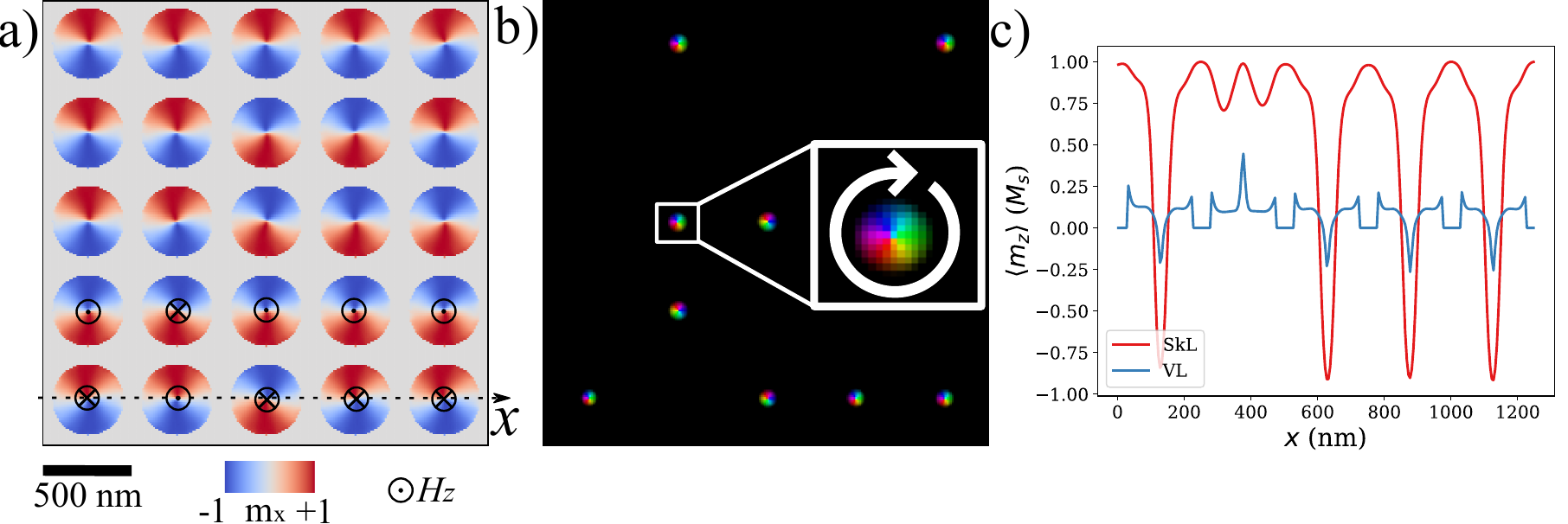}
	\caption{Snapshots of the magnetization  states at $\mu_0H_z = \SI{96}{mT}$ (a) in the vortex layer, showing the $x$ component of the magnetization, (b) the hue-saturation-value (\texttt{hsv}) magnetic induction map calculated from the magnetization state in the skyrmionic layer showing Bloch skyrmions of different chirality, where the inset shows the direction of the magnetization. We assume full saturation, and the hue is given by $\arctan(m_y/m_x)$ being a measure for the in-plane rotation fo the magnetization. In (c) we show the magnetization along the black arrow depicted in (a), where one can see that a skyrmion is only stable if the magnetic vortex above has a polarity opposed to the applied magnetic field, leading to a coupling between the magnetic vortex and the bloch skyrmion.}
	\label{fig:fig03}
\end{figure*}

A linearly increasing external magnetic field is applied along the out-of-plane (OOP) direction, which rises to $\mu_0H_{z}=\SI{100}{mT}$ in \SI{1000}{ns}.
Figure~\ref{fig:fig03}a) shows snapshots of the magnetization states obtained at $\mu_0H_{z} = \SI{96}{mT}$ in the VL. One can clearly see that both chiralities are present.
At magnetic fields above \SI{40}{mT}, skyrmions start to form in SkL under the magnetic vortices. For the 25 vortices shown here, one obtains only nine skyrmions in the SkL, with the chirality of the skyrmion following the one of the vortex. This is due to the coupling between the vortices, and skyrmions, which can be seen best in Fig.~\ref{fig:fig03}c), where we plot the $m_z$ component of the magnetization sampled along the arrow shown in Fig.~\ref{fig:fig03}a).

Comparing Figs.~\ref{fig:fig03}b and c, one can see that at $x = \SI{400}{nm}$ corresponding to second unit element in the bottom row, that the magnetization of SkL stays rather uniform, and parallel to the applied field, while the vortex core is pointing up.
The remaining positions in this row show a clear negative $m_z$ component, and negative vortex core polarity.

Note that if one uses the stable skyrmion lattice as an input in micromagnetic simulations of a SkL thin film at equal external fields, i.e. without the nanodisk lattice, the skyrmions are annihilated due to the strong magnetic field. Hence, the core coupling leads to an increased stability of the skyrmions. However, one could lower the external fields, and increase the size of the skyrmion in the SkL, but then decouple it, if desired.

Additionally, we investigate the tunability of the skyrmion generation and the manipulation of its properties. Here, we initialize the micromagnetic simulations giving all the vortices in VL the same chirality (counter-clock-wise) and core polarity (-1), and repeat the simulation method explained above. This procedure can be realized by experimental meanings due to different Curie temperatures of the two layers($T^{VL}_{C} \approx \SI{800}{K}$, and $T^{SkL}_{C} \approx \SI{400}{K}$).

\begin{figure*}
	\centering
	\includegraphics[width=\textwidth]{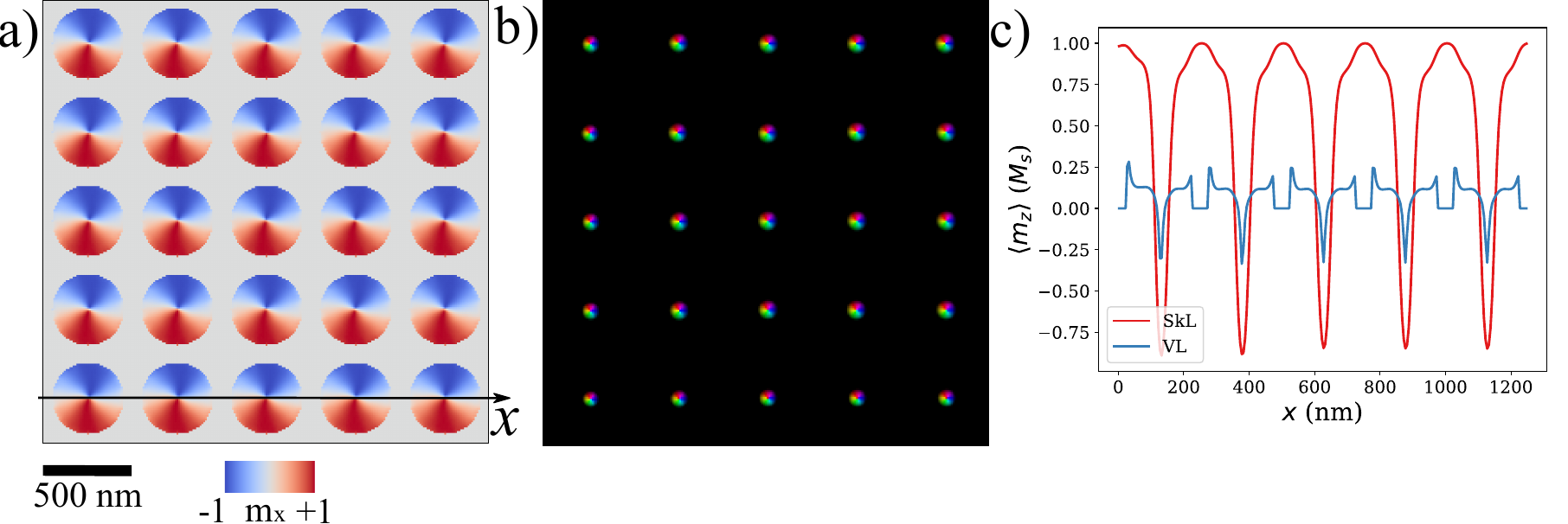}
	\caption{Same as Fig.~\ref{fig:fig03} but the initial states in vortex layer is set manually to have the polarization against the applied OOP field, and a counter-clockwise chirality, while the skyrmion layer is initialized with a random magnetization state. The stable vortex lattice is shown in (a), and the full lattice of Bloch skyrmions with same chirality in (b) obtained at $\mu_0H_z = \SI{96}{mT}$.}
	\label{fig:fig04}
\end{figure*}

Figure~\ref{fig:fig04}a) shows the $m_x$ component of the magnetization, where one can see clearly that all vortices have same chirality. Underneath, in the SkL, we are able to stabilize a full lattice of skyrmions, due their coupling to the vortices, as depicted perpendicularin Fig.~\ref{fig:fig04}b. The local sampling of the magnetization in Fig.~\ref{fig:fig04}c) results in a very periodic modulation of the $m_z$ component based on the presence and equidistant arrangement of the Bloch skyrmions. To increase the visibility of the vortex core polarization, we average $m_z$ three cells along $y$, and sample along $x$.

\subsection{Antiskyrmions in square lattice}
\label{sec:resultsASKL}

Magnetic anti-vortex states have been investigated thoroughly through micromagnetic simulations, as well as experimental works. The magnetization vector field of an antivortex, as stabilized by Gliga and co-authors \cite{gliga_switching_2008}, resembles the \textit{Type-I} configuration in a square ASI lattice, and also the spin-configuration of an antiskyrmion. Hence, we use an interconnected square lattice to stabilized them, and investigate their role as meta-material on the skyrmionic layer.

\begin{figure*}
    \centering
    \includegraphics[width=\textwidth]{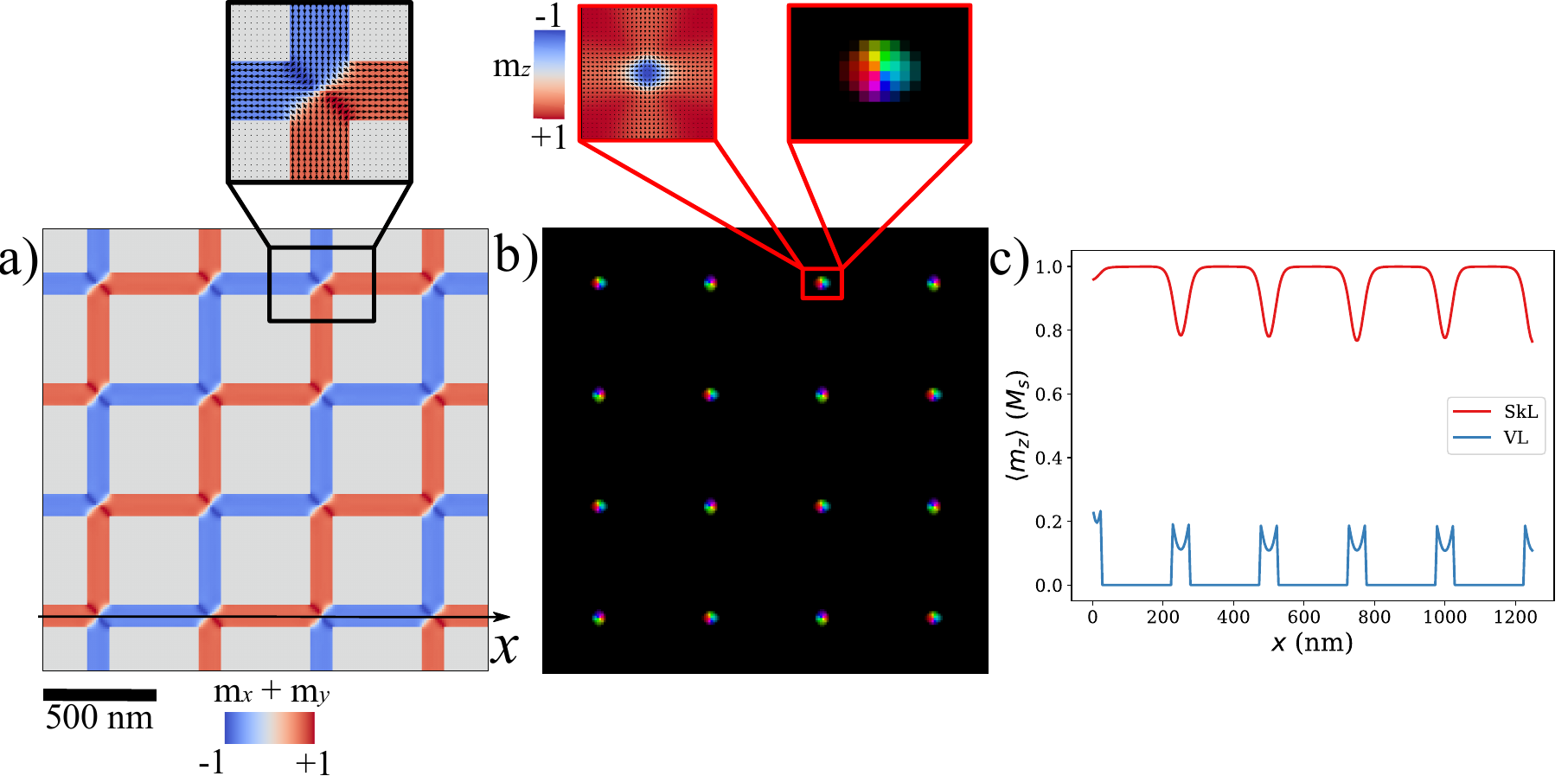}
    \caption{Square lattice where the initial state is set in the antiferromagnetic ground-state according to the \textit{ice-rule}  depicted by the projection of the magnetization along [110] direction in (a), where the zoomed figure helps to see that at the vertex centers of VL antivortices are formed. In the SkL the initial magnetization state is chosen to be random for each cell. The antiskyrmions, shown in (b), are forming for OOP fields large enough to stabilze them, but low enough to not destroy the anti-vortex states. Here, $\mu_0H_z = \SI{97}{mT}$. The zoomed figures are the $m_z$-component of the magnetization (left) with black arrows showing the in-plane components, to better visualize the antiskyrmion states, and the induction-map-like projection fo the magnetization, which depicts the in-plane components of the magnetization as in previous figures.}
    \label{fig:fig05}
\end{figure*}

Figure \ref{fig:fig05}a) depicts the normalized projection of the magnetization along the [110] direction, which is then basically given by the sum $m_x + m_y$ of the magnetization components. Due to presence of the alternating ground state magnetization states (both \textit{Type-I} vertices), the lattice elements are magnetized in an antiferromagnetic manner, leading to the colored pattern presented in Fig.~\ref{fig:fig05}a). With the small inset, and the black arrows, we show how the magnetization state looks at the vertex centers of the lattice, where antivortices are stablized upon relaxation. These remain stable after an OOP field about $\mu_0H_{z}=\SI{100}{mT}$ is applied globally in order to stabilize the spin-textures in SkL.

First, at lower fields, the stray fields generated in the \textit{missing} squares of the lattice, are building a closed loop. This the reason the long-range ordered ground-state in the square ice. The perfectly closed loop of demagnetization fields generates a Bloch skyrmion in the SkL at lower magnetic fields.
Since these new spin-textures are not  coupled, they are not as stable as the skyrmions stabilized by the nanodisk lattice.
However, once a critical field is achieved, the antiskyrmions are stabilized in the SkL, and remain stable as long as the antivortex core does not vanish.

Figure~\ref{fig:fig05}b) shows the stabilized antiksyrmion lattice. The insets illustrate both the $m_z$ component of the magnetization with the black arrows illustrating the in-plane components, and the induction-map simulation, obtained as in previous figures for skyrmions.
The iteration of left and right handed N\'{e}el and Bloch walls can be seen in the $m_z$ snapshot.

Once more, the presence of an antiskyrmion is \textit{solely} dictated by the existence and stability of antivortex core. Figure~\ref{fig:fig05}c) shows the sampled magnetization component $m_z$ along the black arrow from Fig.~\ref{fig:fig05}a), where the core coupling is clearly visible.
Note that the orientation of the nanocross or the chosen \textit{Type-I} configuration will also alter the rotational orientation of the antiskyrmions. 

\subsection{Field pulse driven skyrmion and antiskyrmion annihilation}
\label{sec:resultsDEL}

Even though the idea and visionary examples of the skyrmionic race track memory are very promising and exciting, the large skyrmion sizes, the challenges in their controlled creation, as well as the large bit error rate that the \textit{distance-based} storage approach demonstrates, are major obstacles in the realization of a real  device. Additionally, the precise annihilation of a skyrmion and antiskyrmion will come into play, when one wants to use two distinct spin-textures for the storage of information. An important aspect is that one usually desires a universal writing/deleting method.

Similar to the idea of the exchange spring media~\cite{suess_exchange_2005}, one can make use of hybrid structures in order to facilitate the writing and deleting of spin-objects, while keeping, or even improving the stability of the storage texture. Magnetic vortex or anti-vortex cores have been switched in both numerical investigations~\cite{hertel_ultrafast_2007, gliga_switching_2008, gliga_energy_2011}, as well as experimental realizations~\cite{van_waeyenberge_magnetic_2006, curcic_polarization_2008}, e.g. by applying a pulsed magnetic field.

As we have seen in previous sections, the polarity of the magnetic vortex (antivortex) core determines the existence of a skyrmion underneath. Hence, we aim to switch polarity of the vortex core, and ultimately delete the skyrmion underneath.
For this purpose, we are making use of finite-element based simulations for the accurate micromagnetic description of the vortex core reversal. While the finite-difference codes are very well suitable to describe the static magnetization states, we want to avoid errors induced in the dynamic simulations caused by finite difference discretization of the nanodisks. We use \textit{magnum.fe} to relax an isolated heterostructure \textit{unit-cell} from the nanodisk, and nanocross lattices. By applying the same procedure as above, we are able to stabilize a single skyrmion (anti-skyrmion) coupled to a vortex (anti-vortex).

Similar to well studied magnetic vortex core reversals~\cite{hertel_ultrafast_2007,gliga_switching_2008, gliga_energy_2011} , we apply a pulse of in-plane magnetic field with a maximal amplitude of $\mu_0H^{\mathrm{pulse}}_y = \SI{50}{mT}$ , and a full-width at half maximum of $\sigma_{\mathrm{pulse}} = \SI{100}{ps}$. During our investigations we were able to switch the magnetic vortex cores with in-plane fields as low as $\mu_0H_y = \SI{30}{mT}$ while we the antivortices were switched only for $\mu_0H_y = \SI{50}{mT}$. For a better comparison, we show the results where we apply in both cases $\mu_0H^{\mathrm{pulse}}_y = \SI{50}{mT}$.
\begin{figure}
	\centering
	\includegraphics[width=0.9\columnwidth]{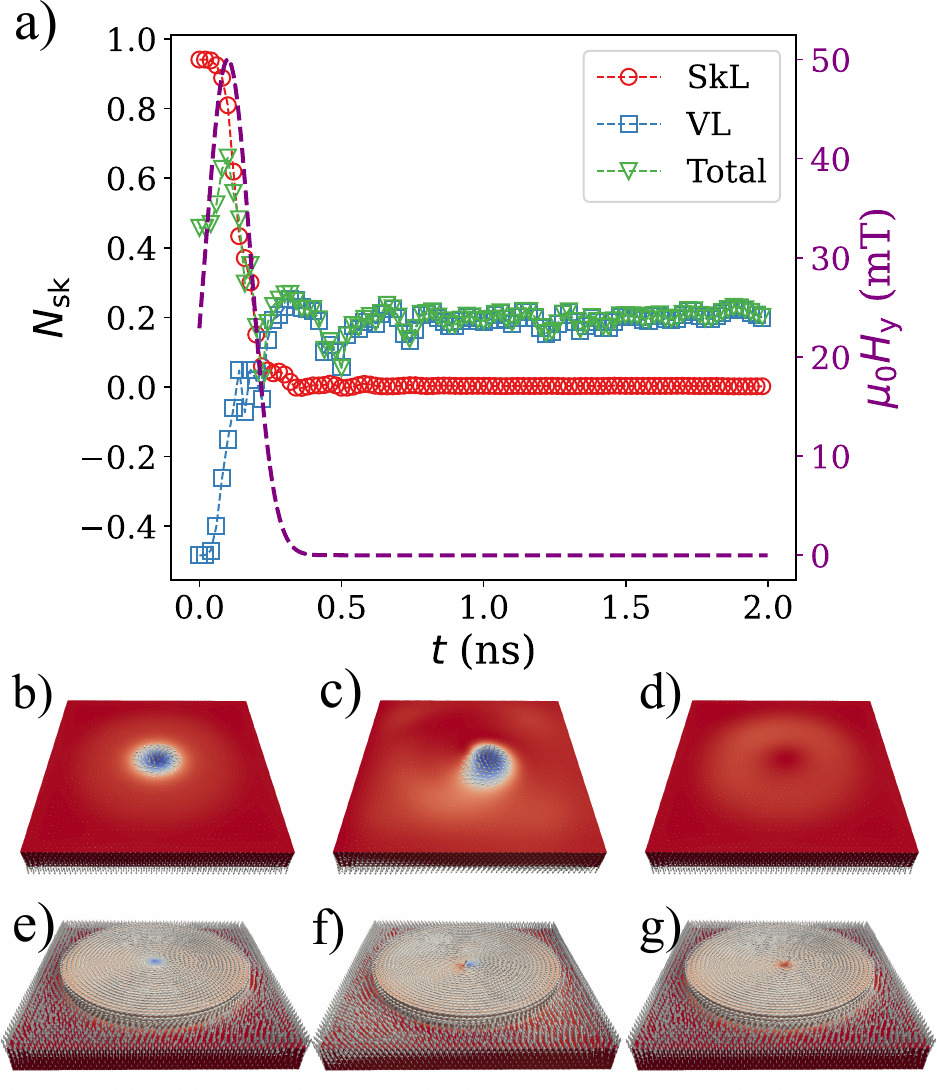}
	\caption{Temporal evolution of the calculated topological charges of the skyrmionic and vortex layers while a pulsed magnetic field is applied in order to reverse the polarity of the vortex core in the vortex layer in (a). The pulsed field is applied with profile shown in purple in (a). The skyrmion is stable at OOP field $\mu_0H_{\mathrm{z}} = \SI{90}{mT}$, as shown in (b), while (e) illustrates the stable magnetic vortex. With the pulsed-field, the vortex-pairs form in (f), which triggers the skyrmion annihilation (c). The skyrmion is fully annihilated (d) once the magnetic vortex cores polarity is reversed (g).}
	\label{fig:fig06}
\end{figure}
\begin{figure}[]
	\centering
	\includegraphics[width=0.9\columnwidth]{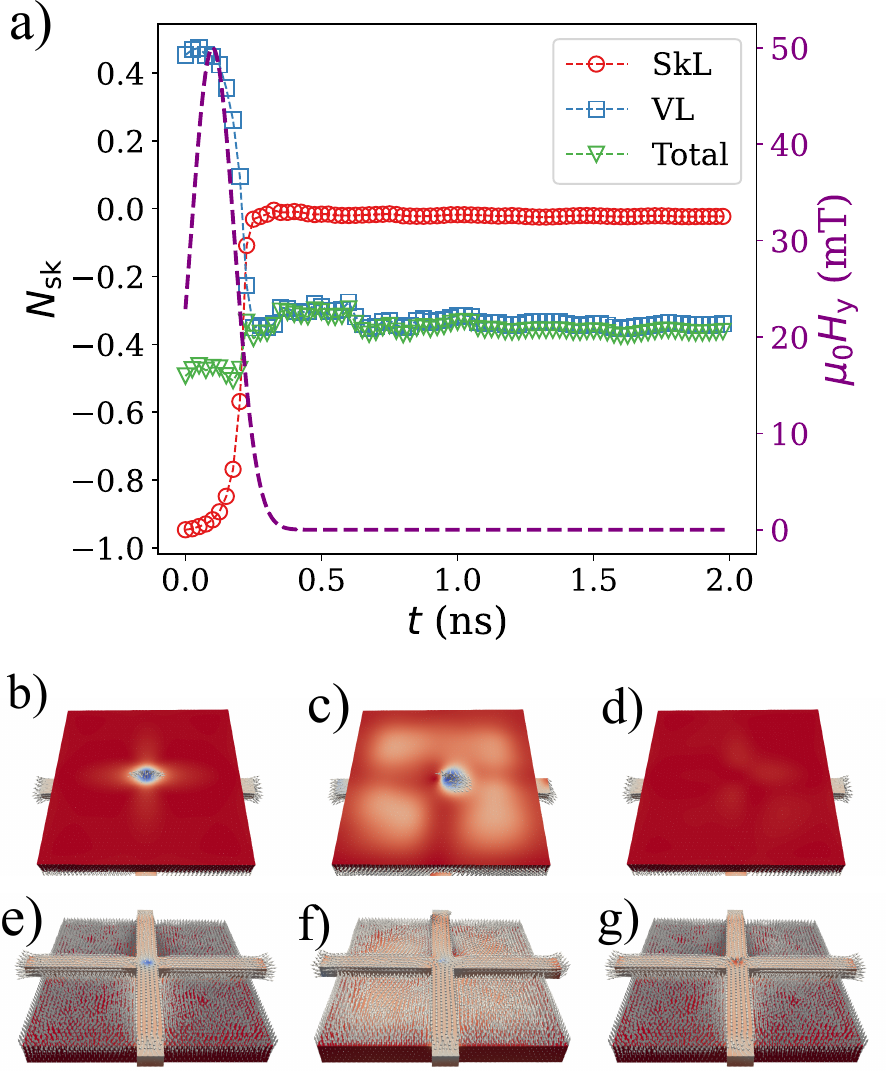}
	\caption{Temporal evolution of the topological charges of the individual layers during the pulsed-field antiskyrmion annihilation process illustrated in (a), with corresponding snapshots of the magnetizations during the annihilation (b) to (g), with the same mechanism as explained in Fig.~\ref{fig:fig06}.}
	\label{fig:fig07}
\end{figure}

Figure~\ref{fig:fig06}a) shows the temporal evolution of the topological number $N_{sk}$ for both the skyrmion and vortex core. Additionally, the applied field pulse profile is plotted along the second $y$-axis. Fig.~\ref{fig:fig06}b)-g) illustrate the magnetization states coressponding the the topological annihilation depicted in Fig.~\ref{fig:fig06}a). One can clearly see that the polarity of the vortex core switches, as expected, over a vortex-antivortex pair formation, which deletes the skyrmion underneath.

In the same manner, we demonstrate the first controlled antiskyrmion annihilation with a pulse of external magnetic field, as described by Fig.~\ref{fig:fig07}. Here, the switching process is not as homogeneous as the skyrmion annihilation, but the antivortex reversal still annihilates the antiskyrmion below. 
Note that both annihilation processes should be reversible in principle, such that one can delete and annihilate skyrmions, as well as antiskyrmions on demand with the same protocol. Both annihilation processes occur at subnanosecond time scales, allowing for a ultrafast deleting (and writing) of promising magnetic storage objects.

\section{Conclusion}
\label{sec:conclusion}
In this paper we make use of micromagnetic simulation to describe the vortex-skyrmion core coupling in magnetic heterostructures consisting of a host material, the skyrmionic layer, and a meta-material, the vortex layer.
We showed that a lattice of nanodisks can be utilized as the meta-material in order to nucleate and stabilize skyrmions in the layer underneath. The controlability of the magnetic vortices regarding the chirality, allows us to tune the handiness of the meta-stable Bloch skyrmions. Additionally, we can decide over the existence of the skyrmions by alligning the magnetic vortex core anti-parallel to the applied OOP field or annihilate it by reversing the magnetic vortex core, such that it is aligned with the external magnetic field.
Furthermore, we demonstrated that an interconnected square lattice, similar to an artificial spin ice, can be used to stabilize antivortices in the vortex layer, which leads to the stabilization of antiskyrmions in the skymrionic layer. The direction of the nanocrosses can be used to tune the direction of the antiskyrmions.
Both spin objects can be easily annihilated by applying a pulsed IP magnetic field, which reverses the  polarity of the magnetic (anti)vortex core, ultimately leading to the collapse of the (anti)skyrmion.

\centering \textbf{Acknowledgements}\\
The computational results presented have been achieved in part using the Vienna Scientific Cluster (VSC). S.K., C.A., F.B., and D.S. gratefully acknowledge the Austrian Science Fund (FWF) for support through grant No. I 4917 (MagFunc).

\bibliography{manuscript}

\end{document}